\documentclass[aps,prl,twocolumn,numerical,superscriptaddress,nofootinbib,showpacs,showkeys,longbibliography]{revtex4-1}

\usepackage{graphicx,color}
\usepackage{amsmath,amssymb,amsfonts}

\newcommand{\be}{\begin{equation}}
\newcommand{\ee}{\end{equation}}
\newcommand{\ba}{\begin{eqnarray}}
\newcommand{\ea}{\end{eqnarray}}
\newcommand{\ban}{\begin{eqnarray*}}
\newcommand{\ean}{\end{eqnarray*}}

\def\v2{\mbox{$v_2$}}

\def\sqrtsNN{\mbox{$\sqrt{s_{\mathrm{NN}}}$}}

\begin{document}

\title{Scaling properties of background- and  chiral-magnetically-driven charge separation \\ in heavy ion collisions at $\sqrt s_{\mathrm{NN}}=200$~GeV}
\medskip

\author{Roy~A.~Lacey} 
\email{Roy.Lacey@stonybrook.edu}
\affiliation{Depts. of Chemistry \& Physics, Stony Brook University, Stony Brook, New York 11794, USA}

\author{Niseem~Magdy} 
\affiliation{Department of Physics, University of Illinois at Chicago, Chicago, Illinois 60607, USA}
\author{Petr~Parfenov}
\affiliation{National Research Nuclear University MEPhI, Moscow 115409, Russia}
\author{Arkadiy~Taranenko}
\affiliation{National Research Nuclear University MEPhI, Moscow 115409, Russia}

%

%
%
%
%

\date{\today}

\begin{abstract}
The Anomalous Viscous Fluid Dynamics model, AVFD,  is used in concert with the charge-sensitive correlator $R_{\Psi_2}(\Delta S)$ to investigate the scaling properties of background- and chiral-magnetically-driven (CME) charge separation ($\Delta S$), characterized by the inverse variance $\mathrm{\sigma^{-2}_{R_{\Psi_2}}}$ of the $R_{\Psi_{2}}(\Delta S)$ distributions obtained in collisions at $\sqrt s_{\mathrm{NN}}=200$~GeV. The $\mathrm{\sigma^{-2}_{R_{\Psi_2}}}$ values for the background are observed to be event-shape-independent. However, they scale with the reciprocal charged-particle multiplicity $(1/\left<  N_{\rm ch} \right>)$, indicating an essential constraint for discerning background from the signal and a robust estimate of the difference between the backgrounds in Ru+Ru and Zr+Zr collisions. By contrast, the $\mathrm{\sigma^{-2}_{R_{\Psi_2}}}$values for signal + background show characteristic $1/\left<  N_{\rm ch} \right>$  scaling violations that characterize the CME-driven contributions. Corrections to recent ${R_{\Psi_2}(\Delta S)}$ measurements~\cite{STAR:2021mii} that account for the background difference in Ru+Ru and Zr+Zr collisions indicate a charge separation difference compatible with the CME. The results further suggest that 
$\mathrm{\sigma^{-2}_{R_{\Psi_2}}}$ measurements for peripheral and central collisions in concert with $1/\left<  N_{\rm ch} \right>$ scaling, provides a robust constraint to quantify the background and aid characterization of the CME.
\end{abstract}

\maketitle

Ion-ion collisions at the Relativistic Heavy Ion Collider (RHIC) and the Large Hadron Collider (LHC) 
lead to the production of a magnetized chiral relativistic quark-gluon 
plasma (QGP) \cite{Kharzeev:2004ey,Liao:2014ava,Miransky:2015ava,Huang:2015oca,Kharzeev:2015znc}, 
akin to the primordial plasma produced in the early Universe \cite{Rogachevskii:2017uyc,Gorbunov:2011zz} and 
several degenerate forms of matter found in compact stars \cite{Weber:2004kj}. Pseudo-relativistic analogs 
include Dirac and Weyl semimetals \cite{Vafek:2013mpa,Burkov:2015hba,Gorbar:2017lnp}. 
 The study of anomalous transport in the QGP can give fundamental insight not only 
on the complex interplay of chiral symmetry restoration, axial anomaly and gluon 
topology\cite{Moore:2010jd,Mace:2016svc,Liao:2010nv,Kharzeev:2015znc,Skokov:2016yrj}, but also on 
the evolution of magnetic fields in the early Universe \cite{Joyce:1997uy,Tashiro:2012mf}. 

A major anomalous process predicted to occur in the magnetized QGP is the 
chiral magnetic effect (CME) \cite{Fukushima:2008xe}.
It is characterized  by the vector current:
\begin{equation}
\vec{J}_V = \frac{N_{c}e\vec{B}}{2\pi^2}\mu_A, {\rm for}\, \mu_A\neq 0,
\end{equation}
where $N_c$ is the color factor,  $\vec{B}$ is the magnetic field and $\mu_A$ is the axial chemical potential 
that quantifies the axial charge asymmetry or imbalance between right- and left-handed quarks in the 
plasma \cite{Fukushima:2008xe,Son:2009tf,Zakharov:2012vv,Fukushima:2012vr}. 
Experimentally, the CME manifests as the separation of electrical charges along the 
$\vec{B}$-field~\cite{Kharzeev:2004ey,Fukushima:2008xe}. This stems from the fact that 
the CME preferentially drives charged particles, 
originating from the same ``P-odd domain'', along or opposite to the $\vec{B}$-field 
depending on their charge. 

The charge separation can be quantified via measurements of the first $P$-odd 
sine term ${a_{1}}$, in the Fourier decomposition of the charged-particle azimuthal 
distribution~\cite{Voloshin:2004vk}:
\begin{eqnarray}\label{eq:a1}
{\frac{dN_{\rm ch}}{d\phi} \propto 1 + 2\sum_{n} (v_{n} \cos(n \Delta\phi) + a_n \sin(n \Delta\phi)  + ...)}\
\end{eqnarray}
where $\mathrm{\Delta\phi = \phi -\Psi_{RP}}$ gives the particle azimuthal angle
with respect to the reaction plane (${\rm RP}$) angle, and ${v_{n}}$ and ${a_{n}}$ denote the
coefficients of the $P$-even and $P$-odd Fourier terms, respectively. 
A direct measurement of the P-odd coefficients  $a_1$, is not possible due to the 
strict global $\cal{P}$ and $\cal{CP}$ symmetry of QCD.
However, their fluctuation and/or variance $\tilde{a}_1= \left<a_1^2 \right>^{1/2}$ can 
be measured with charge-sensitive  correlators such as the 
$\gamma$-correlator~\cite{Voloshin:2004vk}  and the ${R_{\Psi_2}(\Delta S)}$ 
correlator~\cite{Magdy:2017yje,Magdy:2018lwk,Huang:2019vfy,Magdy:2020wiu}.

The $\gamma$-correlator measures charge separation as:
\begin{eqnarray}
\gamma_{\alpha\beta} =& \left\langle \cos\big(\phi_\alpha +
\phi_\beta -2 \Psi_{2}\big) \right\rangle, \nonumber \quad
\Delta\gamma =& \gamma_{\rm OS} - \gamma_{\rm SS},
\label{eq:2}
\end{eqnarray}
where $\Psi_{2}$ is the azimuthal angle of the $2^{\rm nd}$-order event plane which fluctuates about  the ${\rm RP}$, $\phi$ denote the particle azimuthal emission angles, $\alpha,\beta$ denote the electric charge $(+)$ or $(-)$ and SS and OS represent same-sign ($++,\,--$) and opposite-sign ($+\,-$) charges.

The $R_{\Psi_2}(\Delta S)$ correlator~\cite{Magdy:2017yje,Magdy:2018lwk}
measures charge separation relative to $\Psi_2$ via the ratio:
\be
R_{\Psi_2}(\Delta S) = C_{\Psi_2}(\Delta S)/C_{\Psi_2}^{\perp}(\Delta S), 
\label{eq:4}
\ee
where $C_{\Psi_2}(\Delta S)$ and $C_{\Psi_2}^{\perp}(\Delta S)$ are correlation functions that quantify charge separation $\Delta S$, approximately parallel and perpendicular (respectively) to the $\vec{B}$-field. The charge shuffling procedure employed in constructing these correlation functions ensures identical properties for their numerator and denominator, except for the charge-dependent correlations, which are of interest~\cite{Magdy:2017yje,Magdy:2018lwk}; $C_{\Psi_2}(\Delta S)$ measures both CME- and background-driven charge separation while $C_{\Psi_2}^{\perp}(\Delta S)$ measures only background-driven charge separation.  The inverse variance $\sigma^{-2}_{R_{\Psi_{2}}}$ of the $R_{\Psi_2}(\Delta S)$ distributions serves to quantify the charge separation~\cite{Magdy:2017yje,Magdy:2018lwk,Shi:2019wzi}.

A vexing ongoing debate is whether the charge-sensitive ${R_{\Psi_2}(\Delta S)}$ correlator~\cite{Magdy:2017yje,Magdy:2018lwk,Huang:2019vfy,Magdy:2020wiu} shows the requisite response and sensitivity necessary to (i) discern and characterize CME- and background-driven charge separation and (ii) pin down the influence of the background difference in collisions of Ru+Ru and Zr+Zr isobars. The latter is crucial for resolving the ambiguity reported for recent STAR measurements~\cite{STAR:2021mii}   which sought to determine a possible CME-driven charge separation difference for these isobars. Here, we employ the AVFD model~\cite{Shi:2017cpu,Jiang:2016wve} to chart the $R_{\Psi_2}(\Delta S)$ correlators' response to varying degrees of signal and background, primarily in Au+Au collisions, to evaluate its efficacy for detecting and characterizing CME-driven charge separation in the presence of realistic backgrounds. We find characteristic scaling patterns for the background and scaling violations for signal + background that (i) discern between CME- and background-driven charge separation and (ii) allow a robust estimate of the background difference for the Ru+Ru and Zr+Zr isobars. Corrections to recent STAR $R_{\Psi_2}(\Delta S)$ measurements~\cite{STAR:2021mii}, which accounts for this background difference, give results that suggest a CME-driven charge separation that is larger in Ru+Ru than in Zr+Zr collisions.

The AVFD model, which includes realistic estimates for charge-dependent backgrounds such as resonance decays and local charge conservation (LCC) is known to give good representations of the experimentally measured particle yields, spectra, $v_n$, etc~\cite{Proceedings:2017uei}. Thus, it provides an essential benchmark for evaluating the interplay between possible CME- and background-driven charge separation in actual data.
The model simulates charge separation resulting from the combined effects of the CME and the background. An in-depth account of its implementation can be found in Refs.~\cite{Shi:2017cpu} and \cite{Jiang:2016wve}. In brief, the second-generation Event-by-Event version of the model, called E-by-E AVFD,  uses Monte Carlo Glauber initial conditions to simulate the evolution of fermion currents in the QGP, in concert with the bulk fluid evolution implemented in the VISHNU hydrodynamic code~\cite{Shen:2014vra}, followed by a URQMD hadron cascade stage. Background-driven charge-dependent correlations result from LCC on the freeze-out hypersurface and resonance decays. A time-dependent magnetic field $B(\tau) = \frac{B_0}{1+\left(\tau / \tau_B\right)^2}$, acting in concert with a nonzero initial axial charge density $n_5/s$, is used to generate a CME current (embedded in the fluid dynamical equations), leading to a charge separation along the magnetic field. The peak values $B_0$, obtained from event-by-event simulations~\cite{Bloczynski:2012en}, are used with a relatively conservative lifetime $\tau_B=0.6$ fm/c. The initial axial charge density, which results from gluonic topological charge fluctuations, is estimated based on the strong chromo-electromagnetic fields in the early-stage glasma. The present work uses the input scaling parameters for $n_5/s$ and LCC to regulate the magnitude of the CME- and background-driven charge separation.

Simulated AVFD events were generated for varying degrees of signal and background for a broad set of centrality selections in Au+Au and isobar collisions for analysis with the $R_{\Psi_2}(\Delta S)$ correlator. Here, it is noteworthy that the Monte Carlo Glauber parameters employed in the AVFD calculations for the isobars are similar to those used in the centrality calibrations reported in Ref.~\cite{STAR:2021mii}; cross-checks ensured good agreement between the experimental and simulated $N_{\rm ch}$-distributions for both isobars. 

The event selection and cuts mimic those used in the analysis of experimental data~\cite{STAR:2021mii}. Charged particles with transverse momentum $0.2<p_T<2.0$~GeV/$c$ are used to construct $\Psi_{2}$. Each event is subdivided into two sub-events with pseudorapidity $0.1<\eta<1.0$ (E) and $-1.0<\eta<-0.1$ (W) to obtain $\Psi_{2}^{\mathrm{E}}$ and $\Psi_{2}^{\mathrm{W}}$ and their associated centrality-dependent event-plane resolution factors. The $R_{\Psi_{2}}(\Delta S)$ distributions are determined for charged particles with $0.35<p_T<2.0$~GeV/$c$, taking care to use $\Psi_{2}^{\mathrm{W}}$ for particles within the range $0.1<\eta<1.0$ and $\Psi_{2}^{\mathrm{E}}$ for particles within the range $-1.0<\eta<-0.1$ to avoid possible self-correlations, as well as to reduce the influence of the charge-dependent non-flow correlations. The resulting distributions are corrected [$R_{\Psi_{2}}(\Delta S^{''})$] to account for the effects of particle-number fluctuations and the event-plane resolution~\cite{Magdy:2017yje}. The sensitivity of $R_{\Psi_{2}}(\Delta S)$ to variations in the elliptic flow ($v_2$) magnitude at a selected centrality, is also studied using event-shape selection via fractional cuts on the distribution of the magnitude of the $q_2$ flow vector \cite{Schukraft:2012ah}; for a given centrality, the magnitude of $v_2$ is increased(decreased) by selecting events with larger(smaller) $q_2$ magnitudes. This analysis aspect is performed with three sub-events ($A[\eta < -0.3]$, $B[|\eta| < 0.3]$, and $C[\eta > 0.3]$) using the procedures outlined earlier and $q_2$ selection in sub-event $B$. 
%
%
\begin{figure}[t]
\includegraphics[width=1.0\linewidth, angle=-00]{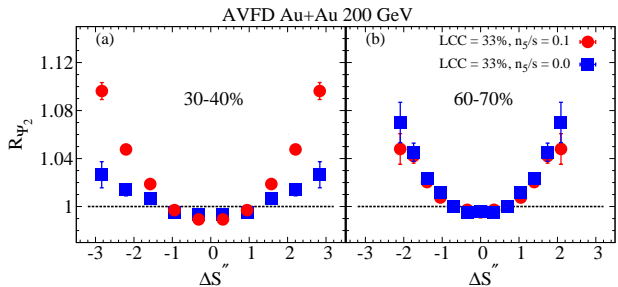}
\vskip -0.050in
\caption{ Comparison of the $R_{\Psi_2}(\Delta S)$ distributions for signal + background (solid circles) and background without signal (solid squares) for 30-40\% (a) and 60-70\% (b) Au+Au collisions at $\sqrt s_{\mathrm{NN}}=200$~GeV. 
}
\label{fig1} 
\end{figure} 

Figure~\ref{fig1} shows a representative comparison of the distributions obtained for signal ({\it Sig.}) + background ({\it Bkg.}) [ $n_{5}/s = 0.1$ and LCC=33\%] and background without signal [ LCC=33\% and $n_{5}/s = 0.0$ ] in 30-40\%  (a) and  60-70\% (b)  central Au+Au collisions. They show the expected concave-shaped distributions for background and signal + background respectively. For the 60-70\%  centrality cut, similar distributions are indicated for background and signal + background, suggesting a loss of sensitivity to the signal in these peripheral collisions. Such a loss will result if the  $\vec{B}$-field is approximately randomly oriented to $\Psi_2$ in these collisions. For the 30-40\%  centrality cut, Fig.~\ref{fig1} (a) shows a narrower distribution for signal + background than for background. This narrowing indicates that the CME signal increases the magnitude of the charge separation beyond the level established by the background; this increase can be quantified via the fraction of the total charge separation attributable to the CME:
\begin{eqnarray}
	%
	f_{\rm CME} =\frac{[{\sigma^{-2}_{R_{\Psi_2}}}(Sig.+Bkg.)-{\sigma^{-2}_{R_{\Psi_2}}}(Bkg.)]}
{[{\sigma^{-2}_{R_{\Psi_2}}}(Sig.+Bkg.)]}, \
\label{eq:5} 
\end{eqnarray}
evaluated with the inverse variance ($\sigma^{-2}_{R_{\Psi_{2}}}$) of the respective distributions. For the 30-40\% central collisions shown in Fig.~\ref{fig1} (a), $f_{\rm CME} \approx 60\%$. This value is a good benchmark of the sensitivity of the $R_{\Psi_{2}}(\Delta S^{''})$ correlator to CME-driven charge separation of this level of signal ($n_5/s =0.1$) in the presence of charge-dependent background (LCC = 33\%) in Au+Au collisions. 

%
%
\begin{figure}[t]
\includegraphics[width=1.0\linewidth, angle=-00]{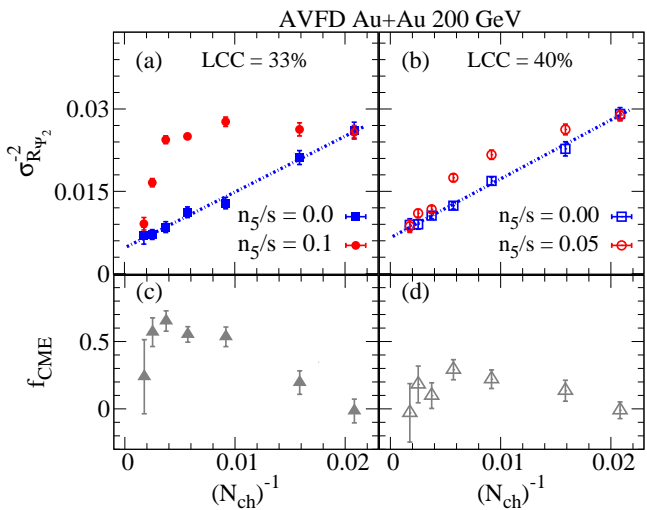}
\vskip -0.10in
\caption{ $\sigma^{-2}_{R_{\Psi_{2}}}$ vs. $1/\left<  N_{\rm ch} \right>$ [(a) and (b)] and $f_{\rm CME}$ vs. $1/\left<  N_{\rm ch} \right>$ [(c) and (d)] for Au+Au collisions at $\sqrt s_{\mathrm{NN}}=200$~GeV,  for two different parameter sets for signal and background as indicated. The dotted lines are drawn to guide the eye. The $f_{\rm CME}$ values in (c) and (d) characterize the fraction of the charge separation which is CME-driven following Eq.~\ref{eq:5}.
} 
\label{fig2} 
\end{figure} 
%

%
%
\begin{figure}[t]
\includegraphics[width=1\linewidth, angle=-00]{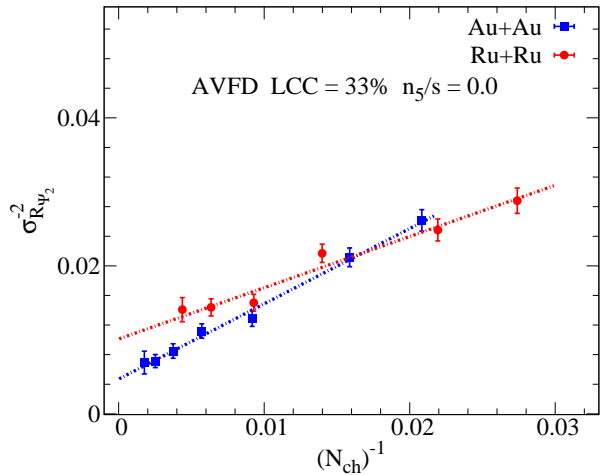}
\vskip -0.10in
\caption{ $\sigma^{-2}_{R_{\Psi_{2}}}$ vs. $1/\left<  N_{\rm ch} \right>$ for the background [LCC=33\% and $n_{5}/s=0.0$] in Au+Au and Ru+Ru collisions at $\sqrt s_{\mathrm{NN}}=200$~GeV.  The dotted lines are drawn to guide the eye.  
} 
\label{fig3} 
\end{figure} 

The centrality dependence of $\sigma^{-2}_{R_{\Psi_{2}}}$ is summarized for Au+Au collisions in Fig.~\ref{fig2} for two different parameter sets for signal and background as indicated. To highlight the scaling property of the background, $\sigma^{-2}_{R_{\Psi_{2}}}$ is plotted vs. $1/\left<  N_{\rm ch} \right>$, where $\left<  N_{\rm ch} \right>$ is the mean number of charged particles employed to evaluate $R_{\Psi_{2}}(\Delta S)$ at the centrality of interest. Figs.~\ref{fig2} (a) and (b)  show that the background scales as  $1/\left<  N_{\rm ch} \right>$, indicating that the observation of this scaling for the experimental $\sigma^{-2}_{R_{\Psi_{2}}}$ measurements would be a strong indication for background-driven charge separation with very little if any, room for a CME contribution. Figs.~\ref{fig2} (a) and (b) also indicate comparable background and signal + background $\sigma^{-2}_{R_{\Psi_{2}}}$ values for large and small $\left<  N_{\rm ch} \right>$. This similarity suggests that background-driven charge separation dominates over the CME-driven contributions in the most central and peripheral collisions. Thus, the $\sigma^{-2}_{R_{\Psi_{2}}}$ measurements for peripheral and central collisions can be leveraged with $1/\left<  N_{\rm ch} \right>$ scaling to give a quantitative estimate of the background over the entire centrality span.

The $\sigma^{-2}_{R_{\Psi_{2}}}$ values, shown for signal + background in Figs.~\ref{fig2} (a), and (b), indicate characteristic positive deviations from the $1/\left<  N_{\rm ch} \right>$ scaling observed for the background. This apparent scaling violation gives a direct signature of the CME-driven contributions to the charge separation. They are quantified with the $f_{\rm CME}$ fractions (cf. Eq.~\ref{eq:5}) shown in Figs.~\ref{fig2} (c) and (d). The indicated $f_{\rm CME}$ values peak in mid-central collisions but reduce to zero at large and small $\left<  N_{\rm ch} \right>$, i.e., central and peripheral collisions. They further indicate that, for these collisions, the $R_{\Psi_{2}}(\Delta S^{''})$  correlator is sensitive to CME-driven charge separation even for a small signal ($n_5/s = 0.05$) in the presence of significant charge-dependent background (LCC = 40\%).

The background $\sigma^{-2}_{R_{\Psi_{2}}}$ values for Au+Au and Ru+Ru collisions are compared in Fig.~\ref{fig3}. The results for Ru+Ru collisions show the same $1/\left<  N_{\rm ch} \right>$ scaling observed for Au+Au. However, they indicate that, for the same centrality, the $\sigma^{-2}_{R_{\Psi_{2}}}$ values for Ru+Ru collisions are larger than those for Au+Au, suggesting a lowering of the sensitivity to the signal in collisions for the isobars.

%
%
\begin{figure}[tb]
\includegraphics[width=1.0\linewidth, angle=-00]{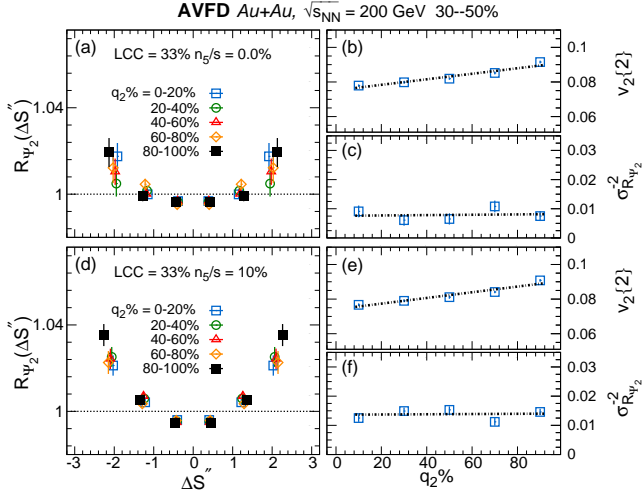}
\vskip -0.10in
\caption{ $q_2$ dependence of the $R_{\Psi_{2}}(\Delta S^{''})$ distributions for background without signal [(a), (b) and (c)] and signal + background [(d), (e) and (f)]. The respective panels show the $q_2$-selected  $R_{\Psi_{2}}(\Delta S^{''})$ distributions [(a) and (d)], the corresponding $v_2$  values [(b) and (e)], and the $\sigma^{-2}_{R_{\Psi_{2}}}$ values [(c) and (f)] extracted from the distributions 
in (a) and (d).
} 
\label{fig4} 
\end{figure} 
The $\sigma^{-2}_{R_{\Psi_{2}}}$ values extracted for background and signal + background at a given centrality, were checked to establish their sensitivity to variations in the magnitude of the anisotropic flow coefficient $v_2$. For this, as discussed earlier, event-shape selection via fractional cuts on the distribution of the magnitude of the $q_2$ flow vector \cite{Schukraft:2012ah} was used. 
Representative results for the sensitivity of $\sigma^{-2}_{R_{\Psi_{2}}}$ to a change in the magnitude of $v_2$ [at a given centrality] are shown in Fig.~\ref{fig4} for background without signal [(a), (b) and (c)] and signal + background [(d), (e) and (f)] for Au+Au collisions. The respective panels show the $q_2$-selected  $R_{\Psi_{2}}(\Delta S^{''})$ distributions [(a) and (d)], the corresponding $v_2$  values [(b) and (e)], and the $\sigma^{-2}_{R_{\Psi_{2}}}$ values [(c) and (f)] extracted from the distributions shown in (a) and (d). They indicate that, while $v_2$ shows a sizable increase with $q_2$ (cf. panels (b) and (e)), the corresponding $\sigma^{-2}_{R_{\Psi_{2}}}$ values (cf. panels (c) and (f)) are insensitive to $q_2$ regardless of background or signal + background. 

Similar patterns of insensitivity have been observed for the $q_2$-selected $\sigma^{-1}_{R_{\Psi_{2}}}$ measurements reported for Ru+Ru and Zr+Zr collisions~\cite{STAR:2021mii}. Notably, the reported insensitivity spans a $\Delta v_2$ range (from low to high $q_2$) much larger than the measured difference between the $v_2$ flow coefficients for the two isobars at a given centrality~\cite{STAR:2021mii}, indicating that the $v_2$ difference between the isobars does not lead to an added difference in their $\sigma^{-2}_{R_{\Psi_{2}}}$ values.
Contributing factors to this insensitivity could stem from (i) an effective $\Delta\eta$ gap between the event-plane and the interest particles that suppresses the charge-dependent non-flow correlations and (ii) the charge shuffling employed in the denominator of the correlation functions that comprise the $R_{\Psi_{2}}(\Delta S^{''})$ correlator~\cite{Magdy:2017yje,Magdy:2018lwk}. The latter eliminates the charge-independent flow correlations and reduces the charge-dependent non-flow correlations.

\begin{figure}[t]
\includegraphics[width=0.75\linewidth, angle=-90]{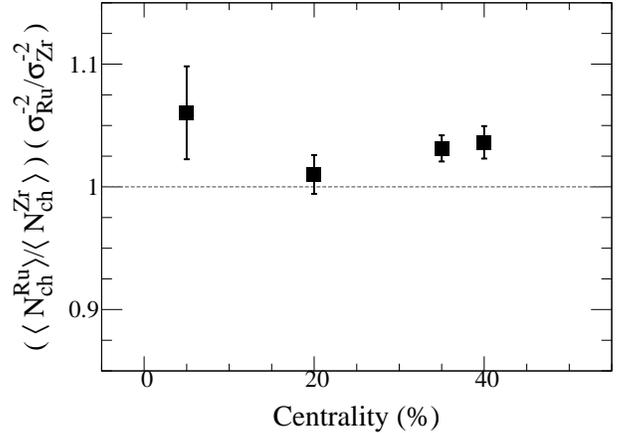}
\vskip -0.10in
\caption{ Centrality dependence of the $\left<  N_{\rm ch} \right>$-scaled ratio of the inverse variance of the $R_{\Psi_{2}}(\Delta S^{''})$ distributions for Ru+Ru (Ru) and Zr+Zr (Zr) collisions at $\sqrt s_{\mathrm{NN}}=200$~GeV. $\left<  N_{\rm ch} \right>$-scaling corrects for the background difference between the two isobars. 
} 
\label{fig5} 
\end{figure}

The ratio of the inverse variance for the two isobars ($\mathrm{\sigma^{-2}_{Ru+Ru}}/\mathrm{\sigma^{-2}_{Zr+Zr}}$)  can also benchmark CME-driven charge separation, which is more prominent in collisions of Ru+Ru than Zr+Zr~\cite{STAR:2021mii}. However, such a ratio must be corrected to account for the background difference between the two isobars. Since $\sigma^{-2}_{R_{\Psi_{2}}}$ is $q_2$-independent and the background scales as $1/\left<  N_{\rm ch} \right>$, a robust estimate for the correction factor at a given centrality is the ratio of the respective  $\left<  N_{\rm ch} \right>$ values for the two isobars. The protocol for the STAR blind-analysis precluded the application of this correction to the $R_{\Psi_{2}}(\Delta S^{''})$ measurements reported in Ref.~\cite{STAR:2021mii}, leading to an ambiguity in the interpretation of measurements that sought to determine a possible CME-driven charge separation difference between the two isobars. Fig.~\ref{fig5} shows the corrected ratios obtained using the $\mathrm{\sigma^{-1}_{Ru+Ru}}/\mathrm{\sigma^{-1}_{Zr+Zr}}$ data reported for several centrality selections in Ref.~\cite{STAR:2021mii}. The $\left<  N_{\rm ch} \right>$-scaled ratios greater than 1.0 are consistent with more significant CME-driven charge separation in Ru+Ru collisions than Zr+Zr collisions.

%
%

In summary, AVFD model simulations that incorporate varying degrees of CME- and background-driven charge separation are used to study the scaling properties of charge separation in heavy ion collisions at $\sqrtsNN~=~200$ GeV. The inverse variance $\mathrm{\sigma^{-2}_{R_{\Psi_2}}}$ of the $R_{\Psi_{2}}(\Delta S)$ distribution, that characterize the charge separation, indicate a linear dependence on $1/\left<  N_{\rm ch} \right>$ which is an essential constraint for discerning background from the signal and a precise estimate of the difference between the backgrounds in Ru+Ru and Zr+Zr collisions. By contrast, the $\mathrm{\sigma^{-2}_{R_{\Psi_2}}}$ values for signal + background show characteristic deviations from the $1/\left<  N_{\rm ch} \right>$  scaling, which serve to characterize the CME-driven contributions to the charge separation. Corrections to recent ${R_{\Psi_2}(\Delta S)}$ measurements~\cite{STAR:2021mii} that account for the background difference in Ru+Ru and Zr+Zr collisions, indicate a charge separation difference between the isobars compatible with the CME. The study further suggest that 
$\mathrm{\sigma^{-2}_{R_{\Psi_2}}}$ measurements for peripheral and central collisions can be leveraged with  $1/\left<  N_{\rm ch} \right>$ scaling to quantify the background and aid characterization of the CME in a wealth of available systems.

\section*{Acknowledgments}
\begin{acknowledgments}
This research is supported by the US Department of Energy, Office of Science, Office of Nuclear Physics, 
under contracts DE-FG02-87ER40331.A008  (RL) and DE-FG02-94ER40865 (NM). 
%
\end{acknowledgments}
%
%
\bibliography{lpvpub} 
\end{document}